\documentclass[preprint,12pt,authoryear]{elsarticle}

\usepackage{amssymb}
\usepackage{amsmath}
\usepackage{color}

\journal{Journal of Computational Physics}

\newcommand {\vect}[1]{\mbox{\boldmath $#1$}}

\newcommand {\dif}[3][]{\frac{d^{#1}#2}{d#3^{#1}}}
\newcommand {\pdif}[3][]{\frac{\partial^{#1}#2}{\partial#3^{#1}}}

\makeatletter
\def\mart{\@ifnextchar[{\mart@@}{\mart@}}
\def\mart@@[#1]#2{\sqrt[#1]{\mathstrut{#2}}}
\def\mart@#1{\sqrt{\mathstrut{#1}}}
\makeatother
\newcommand {\Alfven}{Alfv\'{e}n}

\long\def\symbolfootnote[#1]#2{\begingroup%
\def\thefootnote{\fnsymbol{footnote}}\footnote[#1]{#2}\endgroup}

\newcommand{\apj}{Astrophysical Journal }
\newcommand{\apjs}{Astrophysical Journal Supplement Series}

\newcommand{\pasj}{Publication of the Astronomical Society of Japan}
\newcommand{\aap}{Astronomy and Astrophysics}

\begin{document}

\begin{frontmatter}


 \title{A low-dissipation HLLD approximate Riemann solver for a very wide range of Mach numbers}
\author{Takashi Minoshima\corref{cor1}\fnref{label1}}
\ead{minoshim@jamstec.go.jp}
\ead[url]{https://github.com/minoshim/MLAU}
\author{Takahiro Miyoshi\fnref{label2}}
\cortext[cor1]{Corresponding author}
 \address[label1]{Center for Mathematical Science and Advanced Technology, Japan Agency for Marine-Earth Science and Technology, 3173-25, Syowa-machi, Kanazawaku, Yokohama 236-0001, Japan}
 \address[label2]{Graduate School of Advanced Science and Engineering, Hiroshima University, 1-3-1, Kagamiyama, Higashi-hiroshima 739-8526, Japan}

\begin{abstract}
We propose a new Harten-Lax-van Leer discontinuities (HLLD) approximate Riemann solver to improve the stability of shocks and the accuracy of low-speed flows in multidimensional magnetohydrodynamic (MHD) simulations.
 Stringent benchmark tests verify that the new solver is more robust against a numerical shock instability and is more accurate for low-speed, nearly incompressible flows than the original solver, whereas additional computational costs are quite low.
 The novel ability of the new solver enables us to tackle MHD systems, including both high and low Mach number flows.
\end{abstract}




\end{frontmatter}


\section{Introduction}\label{sec:introduction}
A magnetohydrodynamic (MHD) simulation is an indispensable tool for studying the macroscopic dynamics of laboratory, space, and astrophysical plasmas.
For compressible MHD simulations, shock-capturing schemes have been developed based on the solution to the Riemann problem in one-dimensional hyperbolic conservation laws, which allows us to tackle a situation including supersonic flows.
In particular, the Harten-Lax-van Leer discontinuities (HLLD) approximate Riemann solver developed by \cite{2005JCoPh.208..315M} is extensively implemented in modern MHD simulation codes by virtue of its robustness and accuracy \cite[e.g.,][]{2006A&A...457..371F,2007ApJS..170..228M,2008ApJS..178..137S,2009JCoPh.228..952L,2011PhPl...18b2105Z,2019PASJ...71...83M}.

In practical multidimensional MHD simulations, however, familiar shock-capturing schemes may suffer from numerical difficulties, which include a numerical shock instability for high Mach number flows and a degradation of the solution accuracy for low Mach number flows.
We proposed a multistate low-dissipation advection upstream splitting method \cite[MLAU;][]{2020ApJS..248...12M} to remedy these difficulties, which is an MHD extension of all-speed advection upstream splitting methods for hydrodynamics \citep{2006JCoPh.214..137L,2011AIAAJ..49.1693S,2013JCoPh.245...62K}.
The MLAU scheme provides highly robust and accurate solutions of stringent problems, including high and low Mach number flows, whereas familiar shock-capturing schemes fail to resolve them, and it preserves the MHD discontinuities comparable to the HLLD scheme.
This novel ability allows reliable simulations of wide-ranging Mach number flows in magnetized plasma.
For users of the HLLD scheme to easily enjoy this ability, we propose a new HLLD approximate Riemann solver that implements the techniques used in the MLAU scheme.

\section{Low-dissipation HLLD approximate Riemann solver}\label{sec:new-low-dissipation}
We consider one-dimensional MHD equations written in the following conservative form:
\begin{eqnarray}
&& \pdif{\vect{U}}{t} + \pdif{ \vect{F}}{x} = 0,\label{eq:1}\\
&&\vect{U} = 
\begin{bmatrix}
 \rho, 
 {\rho u}, 
 {\rho v}, 
 {\rho w}, 
 {B_y}, 
 {B_z}, 
 e 
\end{bmatrix}
^T,\label{eq:2}\\
&& \vect{F} = 
\begin{bmatrix}
 \rho u \\
 \rho u^2 + P + |\vect{B}|^2/2-B_x^2 \\
 \rho v u - B_x B_y \\
 \rho w u - B_x B_z \\
 B_y u - B_x v \\
 B_z u - B_x w \\
\left(e+P+|\vect{B}|^2/2\right)u - B_x\left(\vect{u}\cdot\vect{B}\right)\\
\end{bmatrix}
,\label{eq:3}
\end{eqnarray}
where $\vect{U}$ and $\vect{F}$ are the state vector of conservative variables and the corresponding flux vector, respectively, and $\rho,\vect{u}=(u,v,w),\vect{B}=(B_x,B_y,B_z)$, and $e$ are the mass density, velocity, magnetic field, and total energy density.
The gas pressure $P$ is determined from the equation of state for the ideal gas,
\begin{eqnarray}
P = \left(\gamma-1\right)\left(e-\frac{\rho |\vect{u}|^2}{2} - \frac{|\vect{B}|^2}{2}\right),\label{eq:4} 
\end{eqnarray}
where $\gamma$ is the specific heat ratio.
The solenoidal condition of the magnetic field gives $B_x = {\rm constant}$ in one dimension.
Equation (\ref{eq:1}) is discretized on computational cells into a finite volume form as follows:
\begin{eqnarray}
\dif{\vect{U}^n_i}{t} = - \frac{\vect{\hat{F}}_{i+1/2}-\vect{\hat{F}}_{i-1/2}}{\Delta x},\;\;\; \vect{U}^n_i=\frac{1}{\Delta x}\int_{x_i-\Delta x/2}^{x_i+\Delta x/2} \vect{U}(x,t_n)dx,\label{eq:5}
\end{eqnarray}
where $\vect{\hat{F}}_{i \pm 1/2}$ is the numerical flux at the interfaces of a cell $I_i=[x_i-\Delta x/2,x_i+\Delta x/2]$.
The quality of the numerical solutions largely relies on an evaluation of the numerical flux.

The HLLD scheme solves the MHD Riemann problem at a cell interface for the left- and right-side variables $\vect{U}_{L,R}$ as an initial state by allowing five eigenmodes in the Riemann fan.
To obtain four intermediate states in the Riemann fan, the HLLD scheme consistently assumes that the normal velocity $S_M$ and the total (gas + magnetic) pressure $P_t^*$ are constant over the Riemann fan,
\begin{eqnarray}
 S_M &=& \frac{(S_R-u_R)\rho_R u_R - (S_L-u_L)\rho_L u_L - (P_{tR}-P_{tL})}{(S_R-u_R)\rho_R - (S_L-u_L)\rho_L},\label{eq:6}\\
 P_{t}^* &=& \frac{(S_R-u_R)\rho_R P_{tL} - (S_L-u_L)\rho_L P_{tR}+\rho_L\rho_R(S_R-u_R)(S_L-u_L)(u_R-u_L)}{(S_R-u_R)\rho_R - (S_L-u_L)\rho_L},\nonumber \\
\label{eq:7}
\end{eqnarray}
where $P_t = P+|\vect{B}|^2/2$, $S_{L}={\rm min}(0,{\rm min}(u_L,u_R)-c_{f,{\rm max}})$ and $S_{R}={\rm max}(0,{\rm max}(u_L,u_R)+c_{f,{\rm max}})$ \textcolor{black}{are the robust estimations of the minimum and maximum signal speeds taken from Equation (67) in \cite{2005JCoPh.208..315M}}, and $c_{f,{\rm max}} = {\rm max}(c_{f,L},c_{f,R})$ is the maximum fast magnetosonic wave speed,
\begin{eqnarray}
c^2_{f} = \frac{1}{2}\left[\left(c^2_a + a^2\right) + \sqrt{\left(c^2_a + a^2\right)^2-4a^2 c^2_{ax}}\right],c^2_a=\frac{\vect{|B|}^2}{\rho},c^2_{ax}=\frac{B^2_x}{\rho},a^2=\frac{\gamma P}{\rho}.\label{eq:8} 
\end{eqnarray}
Given $S_M$ and $P_t^*$, the intermediate states and the corresponding numerical fluxes are algebraically calculated from the jump conditions across the five waves (see \cite{2005JCoPh.208..315M} for details).
In the following subsections, we modify Equations (\ref{eq:6}) and (\ref{eq:7}) using the techniques applied in the MLAU scheme.

\subsection{Shock detection}\label{sec:shock-detection}
Shock-capturing schemes that can preserve the contact discontinuity, such as the Roe \citep{1988JCoPh..75..400B} and HLLD schemes, tend to suffer from a numerical shock instability when a multidimensional shock is well aligned to the grid spacing, and lead to catastrophic solutions such as odd-even decoupling and the Carbuncle phenomena \citep{1994IJNMF..18..555Q}.
\cite{2000JCoPh.160..623L} and \cite{2003JCoPh.185..342K} argued that the pressure difference term in the interface mass flux (numerical flux of the density) is a possible cause of the instability when the flux is nearly parallel to the shock surface.
The MLAU scheme then introduces a shock-detecting factor $\theta$ to eliminate the pressure difference term only at regions dangerous to the instability.
This technique is implemented into the HLLD scheme through the normal velocity as follows:
\begin{eqnarray}
 S_M &=& \frac{(S_R-u_R)\rho_R u_R - (S_L-u_L)\rho_L u_L - \theta (P_{tR}-P_{tL})}{(S_R-u_R)\rho_R -  (S_L-u_L)\rho_L}, \label{eq:9}\\
\theta &=& {\rm min}\left(1,\frac{-{\rm min}(\Delta u,0)+{c}_{f,{\rm max}}}{-{\rm min}(\Delta v, \Delta w, 0)+{c}_{f,{\rm max}}}\right)^a,\label{eq:10}\\
\Delta u &=& u_{i+1,j,k}-u_{i,j,k},\label{eq:11}\\
\Delta v &=& {\rm min}(v_{i,j,k}-v_{i,j-1,k},v_{i,j+1,k}-v_{i,j,k},\nonumber \\
&& v_{i+1,j,k}-v_{i+1,j-1,k},v_{i+1,j+1,k}-v_{i+1,j,k}),\label{eq:12}\\
\Delta w &=& {\rm min}(w_{i,j,k}-w_{i,j,k-1},w_{i,j,k+1}-w_{i,j,k},\nonumber \\
&& w_{i+1,j,k}-w_{i+1,j,k-1},w_{i+1,j,k+1}-w_{i+1,j,k})\label{eq:13},
\end{eqnarray}
\textcolor{black}{where $a>0$ is a free parameter to make the factor $\theta$ (multiplied by the third term in Equation (\ref{eq:9})) quickly approaching zero when the shock surface is nearly parallel to the $z-x$ or $x-y$ plane. Our numerical tests suggest that $a=1$ is insufficient to suppress the instability and $a \geq 2 $ give reasonable solutions. We empirically use $a=4$.}

\subsection{Pressure correction}\label{sec:pressure-correction}
Familiar one-dimensional shock-capturing schemes (Roe and HLLD) build the numerical flux of the normal momentum to include the velocity difference term with a scale of fast magnetosonic speed, $\sim -\rho c_f (u_R - u_L)/2$, which acts as a diffusion to numerically stabilize the compressible flows.
The term is inappropriate for low-speed flows in multiple dimensions because a finite velocity difference in one direction does not necessarily mean that the flow is compressible (i.e., a rotational flow).
A family of all-speed advection upstream splitting methods for hydrodynamics improves the accuracy of low-speed flows by correcting the velocity difference term with a scale of the convection speed \citep{2006JCoPh.214..137L,2011AIAAJ..49.1693S,2013JCoPh.245...62K}.
 The MLAU scheme adopts this strategy, but uses the scale of a modified fast magnetosonic speed $c_u$ instead of the convection speed in consideration of the inequalities of the MHD eigenmodes,
\begin{eqnarray}
c^2_{u} = \frac{1}{2}\left[\left(c^2_a + |\vect{u}|^2\right) + \sqrt{\left(c^2_a + |\vect{u}|^2\right)^2-4|\vect{u}|^2 c^2_{ax}}\right].\label{eq:14} 
\end{eqnarray}
\textcolor{black}{The HLLD scheme calculates the numerical flux of the normal momentum as $\rho S_M^2 + P_{t}^* - B_x^2$, where the both $\rho S_M^2$ and $P_{t}^*$ terms include the velocity difference.
Whereas the velocity difference term from $\rho S_M^2$ scales with the convection speed (not explicitly shown here), the term from $P_{t}^*$ scales with the fast magnetosonic speed (third term in Equation (\ref{eq:7})), and thus it is corrected as follows:}
\begin{eqnarray}
P_{t}^* &=& \frac{(S_R-u_R)\rho_R P_{tL} - (S_L-u_L)\rho_L P_{tR}+\phi \rho_L\rho_R(S_R-u_R)(S_L-u_L)(u_R-u_L)}{(S_R-u_R)\rho_R - (S_L-u_L)\rho_L}.\nonumber \\
 \label{eq:15}
\end{eqnarray}
The factor $\phi$ multiplied by the third term should satisfy $\propto c_u/c_f \; (c_u/c_f \ll 1)$ to improve the accuracy of low Mach number flows and $1 \; (c_u/c_f \geq 1)$ to reduce to the original scheme at high Mach numbers.
We adopt the function used in \cite{2011AIAAJ..49.1693S},
\begin{eqnarray}
 \phi = \chi (2-\chi), \;\;\; \chi={\rm min}(1,{\rm max}(c_{u,L},c_{u,R})/c_{f,{\rm max}}).\label{eq:16}
\end{eqnarray}
\textcolor{black}{Note that the corrected pressure given by Equation (\ref{eq:15}) does not necessarily reduce to the original one (Equation (\ref{eq:7})) even though the flow is purely one dimensional.}

An asymptotic analysis by \cite{2006JCoPh.214..137L} provides a physical interpretation of Equation (\ref{eq:15}).
In the following, we omit the magnetic field for simplicity.
The physical variables are normalized such that each nondimensional variable remains on the order of unity, i.e., the spatial coordinates by the characteristic length $L$, the density by the ambient density $\rho_0$, the velocity by the reference velocity $u_0$, and the gas pressure by $\rho_0 a_0^2$, where $a_0$ is the characteristic speed of sound.
When one considers compressible flows, the time should be normalized by $L/a_0$, and the resulting momentum equation is as follows:
\begin{eqnarray}
 \pdif{\tilde{\rho} \tilde{\vect{u}}}{\tilde{t}} + M \tilde{\nabla} \cdot (\tilde{\rho} \tilde{\vect{u}} \tilde{\vect{u}}) + \frac{1}{M} \tilde{\nabla} \tilde{P}=0,\label{eq:17} 
\end{eqnarray}
where $M=u_0/a_0$ is the reference Mach number, and the nondimensional variables are denoted by a tilde. Expanding the pressure with respect to $M$, Equation (\ref{eq:17}) implies that for $M \ll 1$,
\begin{eqnarray}
 \tilde{P}=\tilde{P}^{(0)}(\tilde{t})+M \tilde{P}^{(1)}(\tilde{\vect{r}},\tilde{t}).\label{eq:18}
\end{eqnarray}
When one considers nearly incompressible flows, however, it is appropriate to normalize the time by $L/u_0$, and the resulting momentum equation is as follows:
\begin{eqnarray}
 \pdif{\tilde{\rho} \tilde{\vect{u}}}{\tilde{t}} + \tilde{\nabla} \cdot (\tilde{\rho} \tilde{\vect{u}} \tilde{\vect{u}}) + \frac{1}{M^2} \tilde{\nabla} \tilde{P}=0,\label{eq:19}
\end{eqnarray}
which implies that the pressure is constant in space up to the first order in $M$ for $M \ll 1$,
\begin{eqnarray}
 \tilde{P}=\tilde{P}^{(0)}(\tilde{t})+M^2 \tilde{P}^{(2)}(\tilde{\vect{r}},\tilde{t}).\label{eq:20}
\end{eqnarray}
Owing to the difference between Equations (\ref{eq:18}) and (\ref{eq:20}), compressible fluid simulations may overestimate the spatial fluctuation of pressure in low Mach number flows, and their solution deviates from a correct solution with decreasing Mach number.

To cover an incompressible range with the compressible scheme, we consider the modified momentum equation:
\begin{eqnarray}
 \pdif{\rho u}{t} + \pdif{}{x}\left( \rho u^2 + \frac{P_t}{\phi} \right)=0,\label{eq:21}
\end{eqnarray}
where $\phi \propto M \; (M \ll 1)$ and $\phi = 1 \; (M \geq 1)$.
\textcolor{black}{By assuming $\phi$ is uniform over the Riemann fan, the normal velocity is evaluated from the HLL average of the modified momentum equation (as is done in the HLLD scheme):}
\begin{eqnarray}
 S_M &=& \frac{(S_R-u_R)\rho_R u_R - (S_L-u_L)\rho_L u_L - (P_{tR}-P_{tL})/\phi}{(S_R-u_R)\rho_R - (S_L-u_L)\rho_L}.\label{eq:22}
\end{eqnarray}
\textcolor{black}{Once the normal velocity is determined, the pressure in the Riemann fan is consistently derived from the modified momentum equation to satisfy the jump condition across the $S_{\alpha}$ wave $(\alpha=L \; {\rm or} \; R)$,}
\begin{eqnarray}
P_t^* &=& P_{t\alpha}+\phi \rho_{\alpha}(S_{\alpha}-u_{\alpha})(S_M-u_{\alpha}),\nonumber \\
&=& \frac{(S_R-u_R)\rho_R P_{tL} - (S_L-u_L)\rho_L P_{tR}+\phi \rho_L\rho_R(S_R-u_R)(S_L-u_L)(u_R-u_L)}{(S_R-u_R)\rho_R - (S_L-u_L)\rho_L},\nonumber \\
\label{eq:23}
\end{eqnarray}
which is identical to Equation (\ref{eq:15}).
The third term in Equation (\ref{eq:22}) is divided by $\phi$, and one needs a cutoff Mach number to avoid division by zero \citep{2006JCoPh.214..137L}.
Even though division by zero is avoided, this term severely restricts the CFL condition of an explicit scheme.
We therefore do not adopt Equation (\ref{eq:22}) to the new scheme for simplicity.

\textcolor{black}{Relationship between the new scheme and the original scheme is discussed in the one-dimension case.
Since the normal velocity is unchanged in one dimension, the only difference is the total pressure in the Riemann fan, which is rewritten as}
\begin{eqnarray}
 P_{t}^* &=& P_{t,{\rm HLLD}}^* + \frac{(\phi-1) \rho_L\rho_R(S_R-u_R)(S_L-u_L)(u_R-u_L)}{(S_R-u_R)\rho_R - (S_L-u_L)\rho_L},\nonumber \nonumber \\
 &\simeq& P_{t,{\rm HLLD}}^* -\frac{(\phi-1)\bar{\rho} c_f \Delta x}{2} \frac{\Delta u}{\Delta x}, \label{eq:25}
\end{eqnarray}
\textcolor{black}{where $\bar{\rho}=2\rho_L \rho_R/(\rho_R+\rho_L)$, $\Delta u = u_R-u_L$, and we approximate $S_R-u_R=u_L-S_L=c_f$ for subsonic flows.
Substituting Equation (\ref{eq:25}) into Equation (\ref{eq:3}) and recalling the fact that the HLLD scheme satisfies the jump conditions to calculate the numerical flux $\vect{F}_{\rm HLLD}=\vect{F}(\vect{U}_{\rm HLLD}^*)$, we obtain}
\begin{eqnarray}
 \vect{F} \simeq \vect{F}_{\rm HLLD}-\frac{(\phi-1)\bar{\rho} c_f \Delta x}{2} \left[0,\frac{\Delta u}{\Delta x},0,0,0,0,\left(1+M\right)\frac{\Delta \left(u^2/2\right)}{\Delta x}\right]^T.\label{eq:26}
\end{eqnarray}
\textcolor{black}{This numerical flux is interpreted as the combination of the HLLD flux and the anti-diffusion terms for the normal momentum and kinetic energy $(\phi-1 \leq 0)$; however, it does no longer satisfy the jump conditions. }

 To summarize, we propose the new HLLD scheme that adopts Equations (\ref{eq:9}) and (\ref{eq:15}) instead of Equations (\ref{eq:6}) and (\ref{eq:7}). 
These modifications do not violate the preservation of the MHD discontinuities inherent in the original scheme. 
Hereafter, this scheme is termed a low-dissipation HLLD (LHLLD) approximate Riemann solver.

\section{Numerical experiments}\label{sec:numer-exper}
Numerical experiments are conducted to assess the capability of the LHLLD scheme.
The design of the numerical code is the same as that used in \cite{2020ApJS..248...12M}; physical variables are interpolated using the second-order MUSCL scheme with a minmod limiter \citep{1979JCoPh..32..101V}, and are integrated in time through the third-order strong stability preserving Runge-Kutta method \citep{1988JCoPh..77..439S}.
\textcolor{black}{We do not use the characteristic decomposition for low-speed flows because it may cause unphysical oscillation similar to the MLAU scheme (see Section 3.4 in \cite{2020ApJS..248...12M} for details).}
The solenoidal condition of the magnetic field is preserved within the machine precision by the central upwind constrained transport method \citep{2019ApJS..242...14M}.
A CFL number of 0.4 is used.

We confirm that solutions with the LHLLD scheme are indistinguishable from those with the HLLD scheme in standard benchmark tests such as one-dimensional shock tube and Orszag-Tang vortex problems; thus, they are not shown here.
\textcolor{black}{This implies that the inconsistency with respect to the jump conditions (Equation (\ref{eq:26})) has little impact on at least these tests.}
Furthermore, the LHLLD scheme is almost identical to the HLLD scheme in the case of strongly magnetized plasma without strong grid-aligned shocks because the factors $\theta$ and $\phi$ are almost in unity throughout the domain.
We then present two stringent problems conducted in \cite{2020ApJS..248...12M} to focus on the difference between the LHLLD and HLLD schemes, which include extremely low and high Mach number flows in weakly magnetized plasma. 

\begin{figure}[htbp]
\centering
\includegraphics[clip,angle=0,scale=0.3]{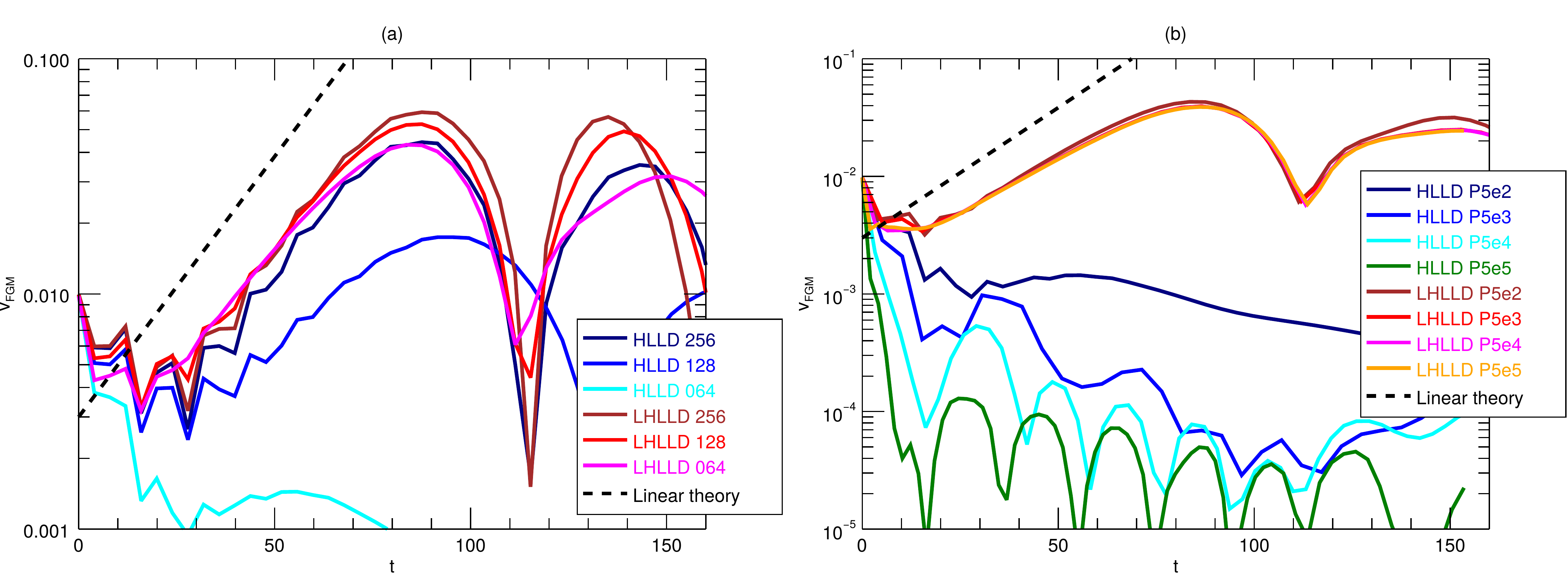}
\caption{Time profile of the Fourier amplitude of the $y$-component of the velocity in the Kelvin-Helmholtz instability. (a) Dependence on grid resolution. (b) Dependence on the initial pressure. The cold- and warm-colored lines are the solutions obtained using the HLLD and LHLLD schemes. The dashed lines indicate the solution obtained with the linear theory.}
\label{fig:khi} 
\end{figure}

The first problem is the two-dimensional Kelvin-Helmholtz instability (KHI) in nearly incompressible flows.
The initial condition has a velocity shear $\vect{u}=((V_0/2)\tanh(y/\lambda),0,0)$, uniform density and pressure $\rho=\rho_0,P=P_0$, and a uniform magnetic field $\vect{B}=B_0(\cos\theta,0,\sin\theta)$, with $\rho_0=V_0=B_0=\lambda=1.0$ and $\theta=71.565^{\circ}$.
We use $\gamma=2.0$.
To initiate the instability, we impose a small $(1\%)$ perturbation into the $y$-component of the velocity around the shear layer with a wavelength of $20 \lambda$ equal to the fastest growing mode.
The computational domain ranging from $0 \leq x<20 \lambda$ and $-10\lambda \leq y<10 \lambda$ is resolved by $N \times N$ cells.
The boundary condition is periodic and symmetric in the $x$ and $y$ directions.
Figure \ref{fig:khi}(a) shows the growth rate obtained using the HLLD and LHLLD schemes at $N=64$, $128$, and $256$ and $P_0=500$.
The linear growth is shown for verification, which is obtained by solving the linearized MHD equations (Equation (68) in \cite{2020ApJS..248...12M}) and is mostly independent of the pressure for nearly incompressible flows. 
The LHLLD scheme converges to the linear growth rate of $\gamma = 0.051V_0/\lambda$ for all runs, whereas the HLLD scheme requires $N=256$ to converge.
 Figure \ref{fig:khi}(b) shows the growth rate at $N=64$ and $P_0=500$, $5000$, $50000$, and $500000$ (corresponding Mach numbers of $1.58\times 10^{-2},\; 5.0\times 10^{-3},\; 1.58\times 10^{-3}$, and $5.0\times 10^{-4}$).
The solutions with the HLLD scheme get worse upon increasing the pressure (decreasing the Mach number), which stems from the fact that the numerical diffusion for velocity scales with a fast magnetosonic speed (Section \ref{sec:pressure-correction}).
By contrast, the solutions with the LHLLD scheme are mostly independent of the Mach number with the help of the pressure correction, although the growth rate and the saturation level at $P_0\geq 5000$ are slightly lower than those at $P_0=500$.

\begin{figure}[htbp]
\centering
\includegraphics[clip,angle=0,scale=0.2]{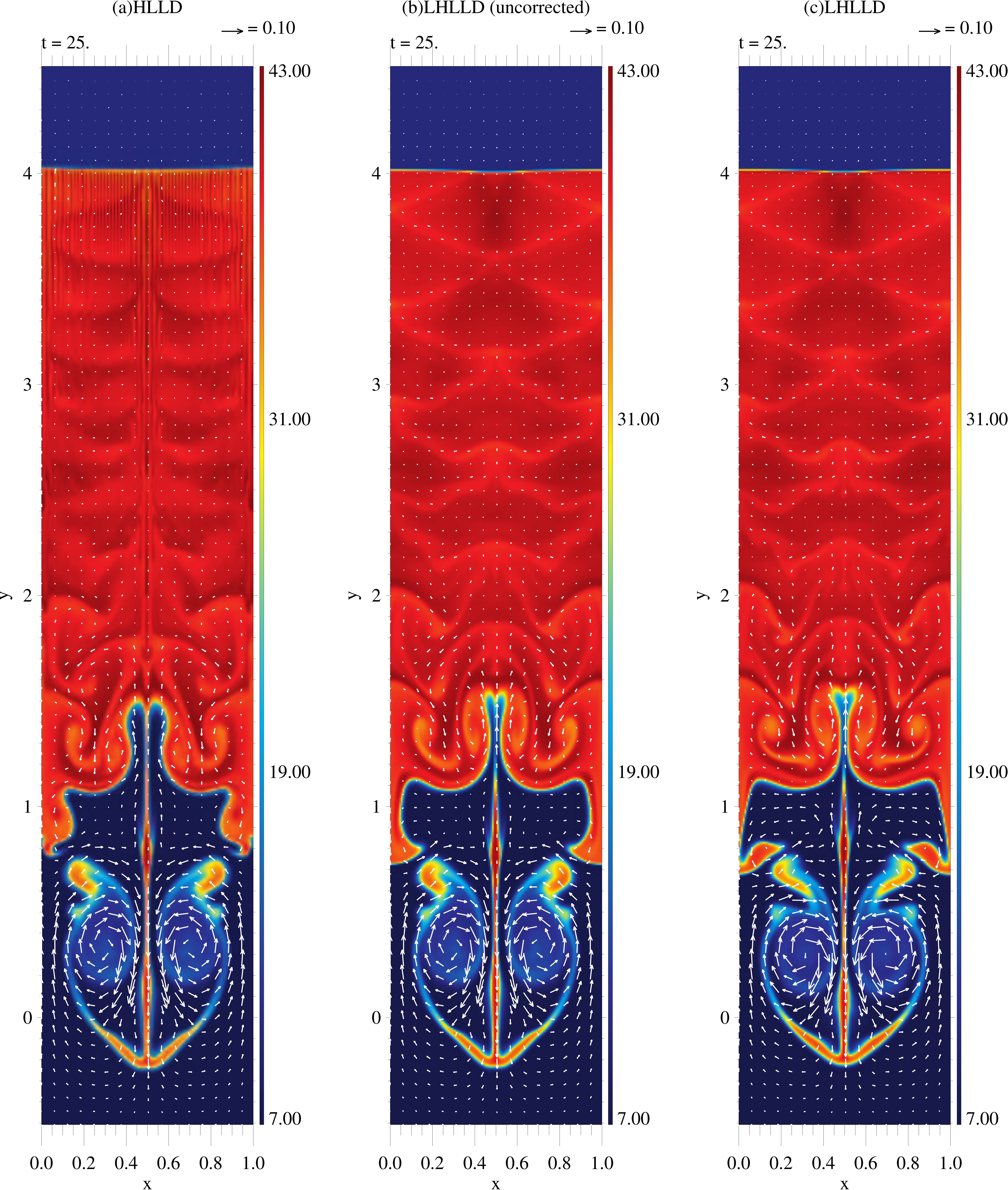}
\caption{Density profile in the Richtmyer-Meshkov instability at $t=25$ obtained using (a) the HLLD scheme, (b) the LHLLD scheme without the pressure correction, and (c) the LHLLD scheme. The arrows represent the rotational velocity component.}
\label{fig:rmi} 
\end{figure}

\begin{figure}[htbp]
\centering
\includegraphics[clip,angle=0,scale=0.3]{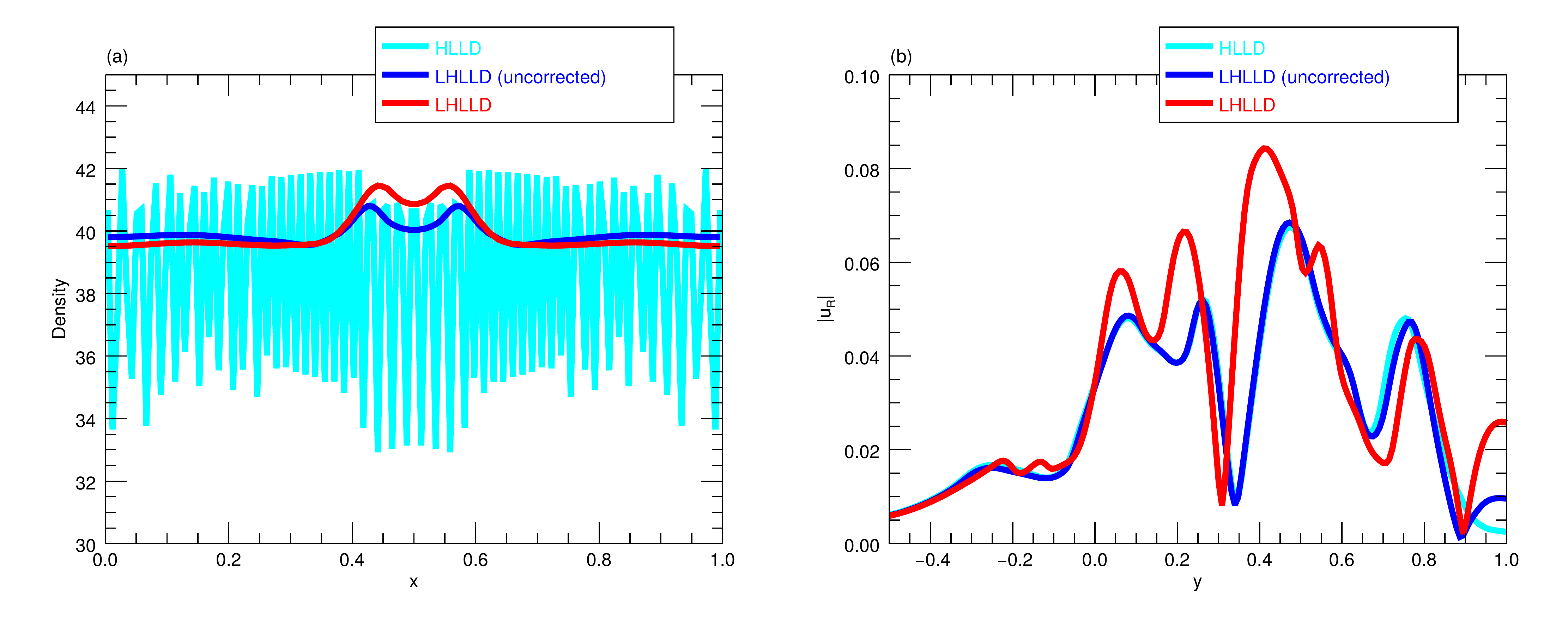}
\caption{One-dimensional profiles of (a) the density along the $x$-direction at $y=4.0$, and (b) the rotational velocity component along the $y$-direction at $x=0.3$. The cyan, blue, and red lines are the solutions obtained using the HLLD scheme, the LHLLD scheme without the pressure correction, and the LHLLD scheme.}
\label{fig:rmi1d} 
\end{figure}

The second problem is the two-dimensional Richtmyer-Meshkov instability (RMI) in hypersonic flows.
The initial condition is as follows:
\begin{eqnarray}
(\rho,\vect{u},\vect{B},P)=
\left\{
\begin{array}{c}
(1.0,0,-1.0,0,0.000034641,0,0,0.00006) \;\;\; (y>0)\\
(3.9988,0,-0.250075,0,0.000138523,0,0,0.75) \;\;\; (y<0)
\end{array}
\right.\label{eq:24}
\end{eqnarray}
which satisfies the Rankine-Hugoniot condition for perpendicular MHD shocks. 
The upstream Mach number is $M=|\vect{u}|/a=100$ and the plasma beta is $\beta=2P/|\vect{B}|^2=10^5$.
We use $\gamma=5/3$.
A corrugated contact discontinuity is imposed in the upstream region, $y_{\rm cd}=Y_{0} + \psi_0 \cos (2 \pi x/\lambda)$, where $Y_{0} = 1.0$, $\psi_0=0.1$ is the corrugation amplitude, and $\lambda=1.0$ is the wavelength.
We shift a frame moving with $v=-0.625$, which is the interface velocity after the collision with the incident shock so that the structure of the RMI remains at approximately $y=0$ throughout the simulation run.
The computational domain ranging from $0 \leq x < \lambda$ and $-40 \lambda \leq y < 40 \lambda$ is resolved by $N \times 80N$ cells, where $N=128$.
The boundary condition is periodic in the $x$ direction and is fixed to the initial state in the $y$ direction.
Figure \ref{fig:rmi} shows the density profile at $t=25$ obtained using (a) the HLLD scheme, (b) the LHLLD scheme without the pressure correction $(\phi=1)$, and (c) the LHLLD scheme.
The grid-scale oscillation observed at the front of the transmitted shock $(y=4)$ in the HLLD scheme is successfully removed with the LHLLD scheme, \textcolor{black}{which is shown in Figure \ref{fig:rmi1d}(a)}.
A comparison between Figure \ref{fig:rmi}(a) and (b) indicates that the oscillation affects the structure until $y=1$, far beyond the shock.
The vortices centered around $(x,y)=(0.3,0.3)$ and $(0.7,0.3)$ are the consequence of the nonlinear development of the RMI, in which the magnetic field is amplified \citep{2012ApJ...758..126S}.
Figure \ref{fig:rmi1d}(b) shows the rotational velocity component $|\vect{u}_R|$ along the $y$-direction $(-0.5<y<1.0)$ at $x=0.3$, which satisfies $\nabla \cdot \vect{u}_R=0$ (Equation (52) in \cite{2019ApJS..242...14M}).
The rotational flow is faster in the run with the LHLLD scheme than in other runs owing to the reduction of the numerical diffusion by the pressure correction, which will affect the saturation level of the magnetic field amplification.
The results of KHI and RMI are in good agreement with those obtained using the MLAU scheme \cite[Sections 4.5 and 4.6 in][]{2020ApJS..248...12M}.

\section{Conclusion} \label{sec:conclusion}
We proposed a new low-dissipation HLLD (LHLLD) approximate Riemann solver that implements the techniques used in the MLAU scheme \citep{2020ApJS..248...12M}.
The LHLLD scheme modifies the normal velocity and the total pressure in the Riemann fan to avoid a numerical shock instability and improve the accuracy of low-speed flows (Equations (\ref{eq:9}) and (\ref{eq:15})).
Stringent benchmark tests verify the capability of the LHLLD scheme.
The scheme is accurate for nearly incompressible flows as long as the flow is super {\Alfven}ic because the velocity difference term in the pressure is corrected to scale with the modified fast magnetosonic speed $c_u \geq c_a$ in consideration of the inequalities of the MHD eigenmodes.
Furthermore, the baseline HLLD scheme intrinsically includes the numerical diffusion for the tangential momentum with a scale of {\Alfven} speed \citep{2020ApJS..248...12M}.
We call this capability as {\it quasi}-all speeds for super-{\Alfven}ic flows.

The clear advantage of the LHLLD scheme over the MLAU scheme is its ease of use for current HLLD scheme users; one only needs to implement two factors $\theta$ and $\phi$ (Equations (\ref{eq:10}) and (\ref{eq:16})).
\textcolor{black}{One can independently implement these factors depending on the purpose, as is demonstrated in Section \ref{sec:numer-exper}. For example, one can use only the pressure correction $\phi$ when the flow is known to be subsonic in a whole domain.}
Additional computational costs are low.
The ratio of the CPU time for the serial computation of a two-dimensional problem in our code is (HLLD:LHLLD:MLAU)=(1:1.03:1.13), although the computational efficiency depends on the design of other procedures (e.g., interpolation) and the level of optimization.
\textcolor{black}{The proposed techniques can be implemented to the HLLC scheme for hydrodynamics \citep{1994ShWav...4...25T} as well, and will be applicable to the simulations with a general equation of state.}
We hope that current HLLD scheme users can easily enjoy the novel ability of the new scheme to tackle MHD systems, including both high and low Mach number flows.

The source code written in C programming language can be downloaded from the GitHub website\footnote{{https://github.com/minoshim/MLAU}}.

\section*{CRediT authorship contribution statement}
{\bf Takashi Minoshima}: Conceptualization, Software, Writing - original draft.
{\bf Takahiro Miyoshi}: Conceptualization, Software, Writing - review \& editing.

\section*{Acknowledgements}
T. Miyoshi is supported by JSPS KAKENHI Grant Numbers JP20K11851,
JP20H00156, JP19H01928.
We would like to thank Editage (www.editage.com) for English language editing.



 \bibliographystyle{elsarticle-harv} 






\gdef\thefigure{\arabic{figure}}


\clearpage

\clearpage

\clearpage

\end{document}